\documentclass[conference,compsoc]{IEEEtran}
\ifCLASSOPTIONcompsoc
  \usepackage[nocompress]{cite}
\else
  \usepackage{cite}
\fi
\ifCLASSINFOpdf
\else
\fi
\newcommand{\etal}{\textit{et al.}}
\usepackage{graphicx}
\usepackage{url}
\begin{document}
%
\title{Towards Model-Driven Dashboard Generation\\ for Systems-of-Systems}

\author{\IEEEauthorblockN{Maria Teresa Rossi}
\IEEEauthorblockA{University of Milano-Bicocca\\
Milan, Italy\\
Email: maria.rossi@unimib.it}
\and
\IEEEauthorblockN{Alessandro Tundo}
\IEEEauthorblockA{University of Milano-Bicocca\\
Milan, Italy\\
Email: alessandro.tundo@unimib.it}
\and
\IEEEauthorblockN{Leonardo Mariani}
\IEEEauthorblockA{University of Milano-Bicocca\\
Milan, Italy\\
Email: leonardo.mariani@unimib.it}}


%


\maketitle

\begin{abstract}

Configuring and evolving dashboards in complex and large-scale Systems-of-Systems (SoS) can be an expensive and cumbersome task due to the many Key Performance Indicators (KPIs) that are usually collected and have to be arranged in a number of visualizations. 
Unfortunately, \emph{setting up} \emph{dashboards} is still a \emph{largely manual and error-prone task requiring extensive human intervention}.

This short paper describes emerging results about the definition of a \emph{model-driven technology-agnostic approach} that can \emph{automatically transform a simple list of KPIs into a dashboard model}, and then translate the model into an actual dashboard for a target dashboard technology. Dashboard customization can be efficiently obtained by \emph{solely modifying the abstract model representation}, freeing operators from expensive interactions with actual dashboards.
\end{abstract}

\begin{IEEEkeywords}
automatic dashboard generation, model-driven engineering, model-based dashboard, systems of systems, monitoring dashboard
\end{IEEEkeywords}

%
\IEEEpeerreviewmaketitle

\section{Introduction}\label{sec:intro}
The proliferation of intelligent software systems and Industrial Internet-of-Things (IIoT) applications (e.g., smart transportation, e-health, or smart grids) demand for the creation of large-scale Cyber-Physical Systems (CPSs) that can cooperate as Systems-of-Systems (SoS)~\cite{Diaz2016,Carlsson2016,Lukkien2016,Morkevicius2017,Fortino2021}.

Monitoring and controlling such SoS is particularly challenging due to the complexity of observing the state of numerous interconnected CPSs that exhibit unpredictable, highly dynamic, and context-dependent behaviors~\cite{Vierhauser2016,Kritzinger2017,Rabiser2018}. Key Performance Indicators (KPIs), such as resource consumption metrics (e.g., CPU and memory usage) and application metrics (e.g., values collected by sensors), are typically computed to monitor and evaluate system behavior and performance. Dashboards have to display these KPIs for the convenient review of the current system state and to gain insights~\cite{Kritzinger2017}. To this end, commercial dashboard tools, such as Grafana~\cite{Grafana} and Kibana~\cite{Kibana}, can be utilized to implement dashboards~\cite{Chakraborty2021}. These tools enable seamless integration with external data sources (e.g., a time series database implemented with Elasticsearch~\cite{ElastichSearchDashboard}) and intuitive customization of dashboard panels.

Despite monitoring systems already offer some degree of automation in terms of configuration (e.g., auto-discovery of monitoring targets~\cite{Calero2015,Trihinas2018,PrometheusSD}, automated deployment of probes~\cite{Tundo2023}, or switchable collection of KPIs~\cite{Trihinas2018,Colombo2022}), the \emph{configuration} of the corresponding \emph{dashboards} is still mainly an \emph{error prone manual activity}~\cite{Tundo2020,Vazquez2020}. In large-scale and complex systems where the set of collected KPIs and monitored services can dynamically and frequently change, maintaining up-to-date dashboards overtime can be a costly and burdensome activity~\cite{Tundo2020,Kintz2017,Vazquez2020}. Consequently, dashboards should be ideally (re)configured based on bare-minimum information (e.g., collected KPIs), and users should rely on dashboard generators for all the visualization-related concerns~\cite{Kintz2017,Tundo2020}.
 
To address this problem, Vázquez-Ingelmo~\etal~\cite{Vazquez2020} presented a model-based approach for generating dashboards that automatically configure themselves through machine learning algorithms.
Erazo-Garzon~\etal~\cite{Erazo2023} proposed a Domain-Specific Language (DSL) and a model-based transformation engine to automatically create dashboards for IoT applications. These studies emphasize the utilization of Model-Driven Engineering (MDE) to manage dashboard evolution. 
In fact, model-driven approaches can employ meta-modeling techniques to create abstract dashboards that adhere to basic guidelines and constraints.

However, state-of-the-art approaches~\cite{Belo2014,Dabbebi2017,Santos2017,Vazquez2020,Tundo2020,DaCol2021,Erazo2023}. suffer from some key limitations. In fact, either operators are asked to \emph{explicitly model the visualizations}, which can be a tedious and time-consuming task, or have \emph{little customization options} once visualizations are generated automatically. Further, existing approaches are all designed to \emph{target specific domains and environments}.

This short paper presents emerging results about the definition of a\emph{model-driven approach} that can \emph{automatically transform a set of user-defined KPIs into a technology-agnostic representation of a dashboard}, which is then \emph{translated into an actual dashboard for a specific target technology}. This chain of transformations provides several benefits: (1) operators can define and generate technology-agnostic dashboards representations, and iterate these activities, regardless of the dashboard technology that is chosen; (2) automatic generation capabilities allow operators with no prior experience with dashboards to quickly obtain useful visualizations; (3) experienced operators can efficiently adapt dashboards modifying the automatically generated models, rather than wasting time interacting with the GUI of dashboard tools; (4) the second automatic transformation is the only step that depends on the selected dashboard technology, and it can be arbitrarily extended to support additional technologies.

\section{Approach}\label{sec:approach}

\begin{figure*}[htb]
    \centering
    \includegraphics[width=.8\textwidth]{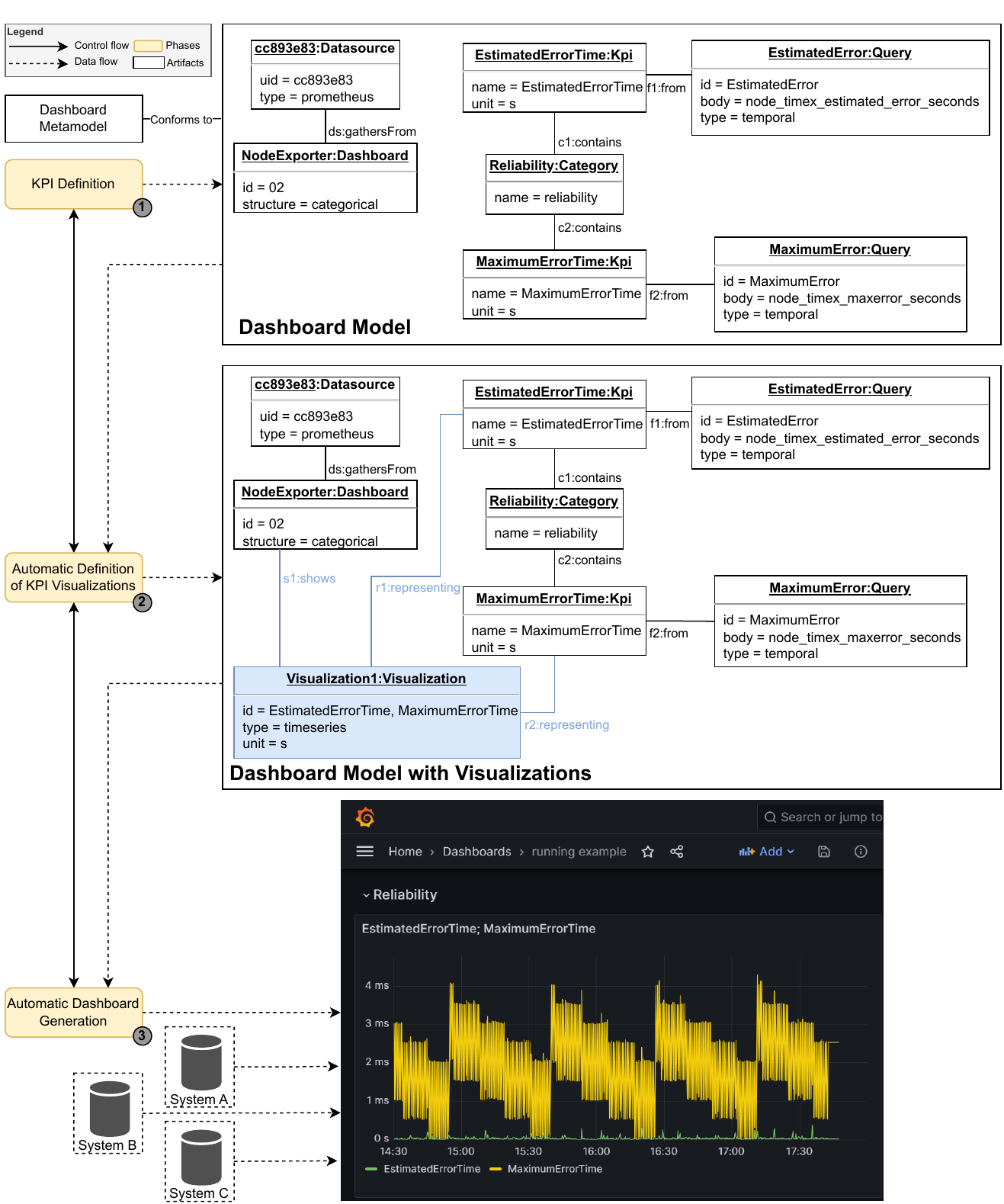}
    \caption{Overview of the Automatic Dashboard Generation Approach.}
    \label{fig:approach}
\end{figure*}

The proposed approach, shown in Figure~\ref{fig:approach} and detailed in this section, consists of three phases, which are represented by the yellow boxes. In the \emph{KPI Definition} phase,
the operator defines the \emph{dashboard model} by specifying a set of KPIs to be visualized in the dashboard by means of a graphical modeling language conforming to a \emph{dashboard metamodel}. Afterwards, the \emph{dashboard model} is utilized by the \emph{Automatic Definition of KPI Visualizations} phase that enriches the model with details about how the KPIs are presented (e.g., how KPIs are distributed among visualizations and the types of visualizations used). Lastly, the \emph{Automatic Dashboards Generation} uses the enriched \emph{dashboard model} to generate the actual dashboard code according to a technology-specific dashboard format (e.g., a Grafana dashboard JSON model). The corresponding artifacts produced by these phases, represented with white boxes, are located on the right-hand side of Figure~\ref{fig:approach}. The solid black line with bidirectional arrows indicates that the operator can freely move through these phases in any order.

\subsection{KPI Definition}
During the \emph{KPI definition} phase, the operator identifies and specifies the KPIs that must be visualized in the dashboard that has to be generated. These KPIs can be collected and stored within any system in a SoS, as long as they are accessible from an interface. In our work, we mainly consider the case of time series databases whose data can be extracted with queries, which is the de facto standard solution to store KPIs in analytics systems. 

The specification produced by the operator must be consistent with a \emph{dashboard metamodel} that specifies all the concepts useful to describe KPIs, including the type of data stored in the KPI and the query that must be executed to extract it.
In particular, we consider KPIs that are organized w.r.t. (i) the categories to which they belong  to (e.g., Reliability, Performance),
(ii) the targets to which they refer to (e.g., a Web server), and (iii) custom structures defined by the user by means of groups of visualizations.

Categories and targets are exploited to automatically define an arrangement of the KPIs and their visualizations in the dashboard (e.g., by grouping the KPIs about a same target or belonging to the same category into a same visualization).
Concerning the KPIs, the operator has to list the KPIs of interest, indicating their names, unit of measures, and the queries to extract them from their data source. When referring to shared ontologies of KPIs (e.g., KPIOnto~\cite{Diamantini2016}), it is only necessary to add the query to extract them.

The output of the KPIs definition phase is a \textit{dashboard model}. In Figure~\ref{fig:approach}, we report an example of a Dashboard that visualizes a set of KPIs collected by the Prometheus Node Exporter~\cite{PrometheusNodeExporter}.
We defined a \textit{Dashboard} that we assume to have a \textit{structure} of type \texttt{categorical}, that is, its visualizations are organized w.r.t. KPI categories. 
To the dashboard we associated a \textit{Datasource} with a unique identifier (i.e., \textit{uid} \texttt{cc893e83}) and of \textit{type} \texttt{prometheus}, that is, the data source used by queries to extract KPI values (i.e., Prometheus~\cite{Prometheus} in this example).
Concerning the KPIs to be visualized, we listed the \texttt{EstimatedErrorTime} and \texttt{MaximumErrorTime} \textit{KPI}s, both using \texttt{s} (i.e., seconds) as \textit{unit} of measure. 
We associated each KPI with a \textit{Query} by defining its \textit{body}. Since we want to monitor  performance over time, we defined every query of \textit{type} \texttt{temporal}, that is, the results will be reported w.r.t. a temporal dimension (e.g., hourly, daily).
Both selected KPIs belong to the \textit{Category} named \texttt{Reliability}.

\subsection{Automatic Definition of KPI Visualizations}
The output model produced in the first phase serves as input for the second phase, which is responsible for determining how to organize and present KPIs in visualizations.
During this phase, visualizations devoted to KPIs representation are \emph{automatically} generated. Specifically, KPIs are associated to visualizations based on their \textit{group} (e.g., \texttt{category}, or \texttt{target}) and \textit{unit} measure. The automatic selection of the visualization \textit{type} (e.g., bar chart, timeseries, gauge) depends on the dimensions (i.e., the \textit{type}) of the KPI queries. For instance, whether there is at least a \texttt{temporal} query associated to a KPI to display, then, the preferred visualization \textit{type} is \texttt{timeseries)}.  
The choice of the groups and visualizations is fully automatic and the operator obtains a model-based representation of an initial dashboard with no additional effort.  

In Figure~\ref{fig:approach}, we report an example of output of the second phase where the initial model is enriched with an instance of a \textit{Visualization} associated to both the \textit{Dashboard} and the included KPIs. Here, the new entity has been generated with an \textit{id} composed by the names of the represented KPIs. As mentioned before, since the KPIs considered in the example use \texttt{temporal} queries, the visualization is of \textit{type} \texttt{timeseries}. In case the generated representation is not fully satisfactory, the operator can manually modify the generated model-based representation of the dashboard.

\subsection{Automatic Dashboards Generation}
The third phase takes as input the dashboard model and produces in output the implementation of the dashboard to be imported into the target dashboard platform. This is the only phase in which a specific dashboard platform is taken into consideration.

To support the generation of the actual dashboard, we use a set of platform-specific translation rules that map visualizations, groups and the other elements in the model into  elements available in  the target platform. For instance, we implemented support for Grafana~\cite{Grafana} and implemented rules to map groups into rows and visualizations into panels, finally obtaining a JSON file that can be directly imported into Grafana. 

On the right-hand side of Figure~\ref{fig:approach}, we show the resulting dashboard that is automatically obtained through a sequence of two transformations starting from the simple list of KPIs to be visualized. 
The row named \textit{Reliability} corresponds to the KPI category, and it contains a visualization with two timeseries: one for the \textit{EstimatedErrorTime} KPI (i.e., green line), and the other for the \textit{MaximumErrorTime} KPI (i.e., yellow line).

\section{Related Work}\label{sec:related}
Previous studies related to our work primarily focus on customizing and automating dashboard creation, which in some cases involve model-based and model-driven methodologies.

Belo~\etal~\cite{Belo2014} and Dabbebi~\etal~\cite{Dabbebi2017} present two distinct methods enabling personalized dashboards to suit users' requirements. Specifically, the former method uses agents to observe users' behavior while consulting dashboards, in order to identify usage patterns and automatically rearrange the dashboards. The latter method presents a conceptual model for developing learning analytics dashboards that can be customized based on the user's usage context and requirements. 
Although useful, both approaches lack of generalization in terms of applicability to different scenarios and classes of users.
In the smart cities scenario, Santos~\etal~\cite{Santos2017} propose an approach for automatically suggesting KPIs to be displayed through the use of knowledge graphs and ontologies. 
However, the user remains responsible for choosing the most appropriate visualizations and customizing the dashboard.

Da Col~\etal~\cite{DaCol2021} propose a machine-learning-based system that assists users in creating data-driven and personalized dashboards by iteratively recommending the most relevant visualization panels. The system can also incorporate user feedback to further customize the suggested panels. 
Deng~\etal~\cite{Deng2023} propose deep reinforcement learning to generate analytical dashboards by exploiting visualization knowledge. Their approach exploits a training environment to let learning agents to explore and imitate human behavior in dashboard generation processes.
Vázquez-Ingelmo~\etal~\cite{Vazquez2020} propose a generator based on the software product lines paradigm that uses machine learning algorithms to infer dashboard features. These collected features are then utilized to create a dashboard metamodel and represent abstract dashboards.
Erazo-Garzon~\etal~\cite{Erazo2023} develop an automated approach for generating web dashboards intended for IoT applications. They introduced a DSL for modeling dashboards and their visualizations. 
Similarly, Kintz~\etal~\cite{Kintz2017} suggest a model-based approach to develop dashboards customized for users within the scope of business process monitoring.
The primary limitations of these previous studies concern their limited applicability and reusability in diverse contexts, and the seamless integration with various data sources, formats, and dashboard technologies.

To partially overcome this limitation, Tundo~\etal~\cite{Tundo2020} propose a declarative approach to create dashboards automatically. They also envision the use of meta-model layouts to determine the visualization style when organizing a set of input KPIs. Although the authors anticipate developing a technology-agnostic solution as future work, the proposed approach concentrates on visual features, and the usage of models is restricted to the visualization templates.

Compared to related work, the aim of our approach is to combine the benefits of MDE to create abstract dashboards and enable non-expert users to handle the complexity of configuring dashboards for SoS, with the benefits of a technology-agnostic solution that can integrate different data sources and target dashboard platforms.

\section{Conclusions and Future Work}\label{sec:conclusions}
The size and unpredictability of the  interactions within SoS necessitate automated solutions that can rapidly reconfigure monitoring dashboards to respond promptly to new and emerging monitoring requirements. Current approaches still limit reusability and impose limitations on the usage of specific environments or technologies.

This short paper describes our ongoing work on the creation of a model-driven technique that can transform a collection of user-specified KPIs automatically into a technology-agnostic and abstract dashboard representation. Subsequently, the dashboard is converted into code for a specific dashboard technology.

Future work primarily involves fully developing the approach to provide a model-driven graphical solution that can automatically generate dashboards for multiple target technologies, requiring users to make minimal effort in defining the KPIs to be displayed, and running use studies to validate the approach.


\ifCLASSOPTIONcompsoc
  \section*{Acknowledgments}
\else
  \section*{Acknowledgment}
\fi

This work has been partially supported by the Centro Nazionale HPC, Big Data e Quantum Computing (PNRR CN1 Spoke 9); the Engineered MachinE Learning-intensive IoT systems (EMELIOT) national research project (PRIN 2020 program Contract 2020W3A5FY); and the COmmunity-Based Organized Littering (COBOL) national research project (PRIN 2022 PNRR program Contract P20224K9EK).



%

\bibliographystyle{IEEEtran}
\bibliography{bib}

\end{document}